\documentclass[aps,twocolumn,groupedaddress,showpacs]{revtex4}

\begin{document}
\bibliographystyle{apsrev}

\title{Fermions, strings, and gauge fields in lattice spin models}

\author{Michael Levin}
\author{Xiao-Gang Wen}
\homepage{http://dao.mit.edu/~wen}
\affiliation{Department of Physics, Massachusetts Institute of Technology,
Cambridge, Massachusetts 02139}

\date{Jan. 2003}

\begin{abstract}
We investigate the general properties of lattice spin models with emerging 
fermionic excitations. We argue that fermions always come in pairs and their 
creation operator always has a string-like structure with the newly created 
particles appearing at the endpoints of the string. The physical implication 
of this structure is that the fermions always couple to a nontrivial gauge 
field. We present exactly soluble examples of this phenomenon in $2$ and $3$ 
dimensions. Our analysis is based on an algebraic formula that relates 
the statistics of a lattice particle to the properties of its hopping 
operators. This approach has the advantage that it works in any number of 
dimensions - unlike the flux-binding picture developed in FQH theory.

\end{abstract}

\maketitle

\section{Introduction}

For many years, it was thought that Fermi statistics were fundamental, in the 
sense that one could only obtain a theory with fermionic excitations by
introducing them by hand (via anti-commuting fields). Then, over the past two
decades, this view began to change. A number of real world and theoretical
examples showed that fermions and anyons could \emph{emerge} as low energy
collective modes of purely bosonic systems. The first examples along these
lines were the fractional quantum Hall states. \cite{TSG8259,L8395} Usually we
think of the FQH states as examples of anyonic excitations emerging from
interacting fermions. However, from a purely theoretical point of view, the
same effects should occur in systems of interacting bosons in a magnetic
field.\cite{ASW8422} There were also indications of emerging fermions in the
slave-boson approach to spin-1/2
systems\cite{BZA8773,BA8880,KL8795,WWZcsp,KW8983,RS9173,Wsrvb} and in the
study of resonating valence bond (RVB)
states.\cite{KRS8765,RK8876,RC8933,MS0181} Unfortunately, the RVB picture and
the slave boson approach both rely on approximate or mean-field techniques to
construct and analyze these exotic states. More recently, a number of
researchers have introduced exactly soluble or quasi-exactly soluble models
with emerging fermions.
\cite{K9721,MS0181,SP0258,BFG0212,MS0204,IFI0203,Wqoexct,M0284} These models
allow for a more well-controlled analysis, albeit in very specific cases. 

The mean-field approach and the exactly-soluble examples both provide clues to
the structure and basic properties of bosonic models with fermionic
excitations. They indicate, among other things, that fermions never appear
alone in lattice spin systems. Instead they always come together with a
nontrivial gauge field and the emerging fermions are 
associated with the deconfined phase of the gauge
field.\cite{WWZcsp,RS9173,Wsrvb} The deconfined phases always seem to contain
a new kind of order - topological order,\cite{Wsrvb,Wtoprev} and the emerging
fermions and anyons are intimately related to the new order. 

This was particularly apparent in the context of the slave-boson approach. In
this technique, one expresses a spin-1/2 Hamiltonian in terms of fermion
fields and gauge fields.\cite{BZA8773,BA8880} Clearly, the presence of fermion
fields does not, by itself, imply the existence of fermionic quasi-particles -
the fermions appear only when the gauge field is in the \emph{deconfined}
phase.  \Ref{KL8795,WWZcsp,KW8983,RS9173,Wsrvb} constructed several deconfined
phases where the fermion fields do describe well defined quasi-particles.
Depending on the properties of the deconfined phases, these quasi-particles
can carry fractional statistics (for the chiral spin states)
\cite{KL8795,WWZcsp,KW8983} or Fermi statistics (for the $Z_2$ deconfined
states).\cite{RS9173,Wsrvb} 

Although it was less evident initially, a similar picture has emerged from the
study of RVB states and the associated quantum dimer models. 
\Ref{KRS8765,RK8876} originally proposed that fermions (spinons) could emerge
from a nearest neighbor dimer model on a square lattice. Later, it was
realized that the fermions couple to a $Z_2$ gauge field, and the fermionic
excitations only appear when the field is in the deconfined
phase.\cite{RC8933} It turns out that the dimer liquid on square lattice with
only nearest neighbor dimers is not in the deconfined phase (except at a
critical point).\cite{RK8876,RC8933} However, on a triangular lattice, the
dimer liquid does have a $Z_2$ deconfined phase.\cite{MS0181} The results in
\Ref{KRS8765,RK8876} are valid in this case and fermionic quasi-particles do
emerge in a dimer liquid on a triangular lattice.

In this paper, we attempt to clarify these observations and to put them on
firmer foundations. We give a general argument which shows that emerging
fermions always occur together with a nontrivial deconfined gauge field. In
addition, we derive an algebraic formula that allows one to calculate
the statistics of a lattice particle from the properties of its hopping
operators. We feel that this formula both elucidates the fundamental meaning
of statistics in a lattice system and simplifies their computation. This
approach has the added advantage that it works in any number of dimensions -
unlike the flux binding picture\cite{ASW8422} developed in FQH theory. We
would like to point out that all the previously mentioned examples of emerging
fermions are two dimensional models. The emerging fermions in those models are
related to the flux binding picture in \Ref{ASW8422}. Our algebraic
approach allows us to establish the emergence of fermions in an exactly 
soluble 3D bosonic model.

Our paper is organized as follows: in the first section, we give a definition
of the statistics of a lattice particle. In the second section, we derive the
algebraic statistics formula discussed above. In the third section, we apply  
the formula to the case of fermionic (or anyonic) excitations in a lattice 
spin system. The formula demands that the fermions always come in pairs, and 
that their pair creation operator has a string-like structure with the newly 
created particles appearing at the ends. We show that the strings represent 
gauge fluctuations, and the fermions carry the corresponding gauge charge. 
After presenting the general argument, we devote the last two sections 
to exactly soluble examples of this phenomenon in $2$ and $3$ dimensions.

The above result suggests a very interesting picture of emerging fermions.
Fermions and gauge fields appear to be two sides of the same coin: in some 
sense, fermions are the ends of strings and gauge fields are the fluctuations 
of strings. Extended string-like structures seem to be the key to 
understanding both Fermi statistics and gauge fields.

\section{Definition of Statistics}

What do we mean when we say that a particle is a ``boson'' or a ``fermion''?
The most common way to define statistics is to use the algebra of creation and 
annihilation operators. If these operators satisfy the bosonic algebra, the 
corresponding particles are said to be bosons; if they satisfy the fermionic 
algebra, the particles are said to be fermions. 

In this paper, we will use another, equivalent, definition based on the path 
integral formulation of quantum mechanics. Given any $m$ particle system with 
short-range interactions, we can write down the corresponding multi-particle 
action $S$. The most general action $S$ has two components:
\begin{displaymath}
S = S_{loc} + S_{top}
\end{displaymath}
The first term, $S_{loc}$, can be any local expression. Typically, it's just 
the usual classical action. The second term, $S_{top}$, is less familiar. In 
general, it can be any expression which only depends on the topology (e.g. 
homotopy class) of paths. In $3$ or more dimensions, the form of $S_{top}$ is 
highly constrained. The amplitude for a closed path $P$ must be of the form 
\begin{displaymath}
e^{iS_{top}} = (\pm 1)^n
\end{displaymath}
where $n$ is the number of particle exchanges that occur in $P$. The statistics
of the particles are defined by the sign in this expression. If this sign is 
positive, we say the particles are``bosons'', if the sign is negative, the 
particles are ``fermions''. In $2$ dimensions, there are other, more 
complicated possibilities for $S_{top}$. These particles are called 
``anyons.''

According to this definition, the statistics are completely determined by the
topological term $S_{top}$ in the action. One way to isolate this term is to
compare the amplitude for two paths which are the same locally (this will be
made more precise in the next section) but differ in their global properties.
In particular, we can compare the amplitude for a path which exchanges two
particles with another path which is the same locally, but doesn't exchange
the two particles. The difference between the phases of the two amplitudes is 
then precisely $e^{iS_{top}} = (\pm 1)$, the statistical phase.

We now reformulate this (theoretical) test of statistics so that it can be
applied to particles on a lattice. The prescription is as follows: we take a 
two particle state $\ket{r_{1}, s_{1}}$, and consider a product of hopping
amplitudes along a lattice path which exchanges the particles
\begin{eqnarray}
\bra{r_{n},s_{n}} H \ket{r_{n-1}, s_{n-1}} \bra{r_{n-1}, s_{n-1}} H 
\ket{r_{n-2}, s_{n-2}} \nonumber \\ 
... \bra{r_{3}, s_{3}} H \ket{r_{2},s_{2}} 
\bra{r_{2}, s_{2}} H \ket{r_{1},s_{1}} 
\label{prod}
\end{eqnarray}
Each hopping amplitude in this expression involves one of the two particles 
moving to a neighboring site while the other particle remains fixed. At the
end, the two particles have exchanged places: $r_{n} = s_{1}, s_{n} = r_{1}$.

Now we construct another path which is the same locally, but doesn't
exchange the two particles, and take a product of hopping amplitudes along
this path. The claim is that the difference between the phases of these 
two expressions is precisely the statistical phase of the particles. That is, 
the particles are bosons, fermions, or anyons depending on whether the 
phase difference is $+1$, $-1$, or something else.

One way to see this is to remember the derivation of the path integral
formulation of quantum mechanics.  According to the standard derivation, the
amplitude $e^{iS}$ for a two particle path $\ket{r(t), s(t)}$ is given by a 
product
\begin{eqnarray*}
e^{iS} \propto \bra{r(t_{n}), s(t_{n})} e^{-iH\Delta t} 
\ket{r(t_{n-1}), s(t_{n-1})} \\ 
...
\bra{r(t_{3}), s(t_{3})} e^{-iH\Delta t} \ket{r(t_{2}), s(t_{2})} \\
\bra{r(t_{2}), s(t_{2})} e^{-iH\Delta t} \ket{r(t_{1}), s(t_{1})}
\end{eqnarray*}
in the limit that $\Delta t = t_{n} - t_{n-1} \rightarrow 0$. 

Now, in the discrete (lattice) case, the above expression can be further 
simplified. Since $\Delta t \rightarrow 0$, we can rewrite it as
\begin{eqnarray*}
e^{iS} \propto \bra{r(t_{n}), s(t_{n})} (1-iH\Delta t) 
\ket{r(t_{n-1}), s(t_{n-1})} \\ 
...
\bra{r(t_{3}), s(t_{3})} (1-iH\Delta t) \ket{r(t_{2}), s(t_{2})} \\
\bra{r(t_{2}), s(t_{2})} (1-iH\Delta t) \ket{r(t_{1}), s(t_{1})}
\end{eqnarray*}
Successive states $\ket{r(t_{i}), s(t_{i})}, \ket{r(t_{i+1}), s(t_{i+1})}$ 
must either be identical, or must differ by a single particle hop. The matrix 
elements between identical states don't contribute to the phase. Thus, we
can drop them without affecting our result. We are left with:
\begin{eqnarray*}
e^{iS} \propto \bra{r(t'_{k}), s(t'_{k})} (-iH\Delta t) 
\ket{r(t'_{k-1}), s(t'_{k-1})} \\ 
...
\bra{r(t'_{3}), s(t'_{3})} (-iH\Delta t) \ket{r(t'_{2}), s(t'_{2})} \\
\bra{r(t'_{2}), s(t'_{2})} (-iH\Delta t) \ket{r(t'_{1}), s(t'_{1})}
\end{eqnarray*}
where the $t'_{k}$s are all distinct. 

Now, as we discussed earlier, the statistical phase can be obtained by 
comparing the phase of this amplitude with another amplitude which is the same 
locally but doesn't exchange the two particles. When we make this comparison, 
the factors of $-i$ drop out (since they contribute equally to the two 
products). Thus, it suffices to compare products of the form:
\begin{eqnarray*}
\bra{r(t'_{k}), s(t'_{k})} H \ket{r(t'_{k-1}), s(t'_{k-1})} \\
...
\bra{r(t'_{3}), s(t'_{3})} H \ket{r(t'_{2}), s(t'_{2})} \\
\bra{r(t'_{2}), s(t'_{2})} H \ket{r(t'_{1}), s(t'_{1})}
\end{eqnarray*}
This is precisely the expression in \Eq{prod}.
\section{Statistics and the hopping operator algebra}

In this section we derive a simple algebraic formula for the statistics of a 
particle hopping on a lattice. This formula is completely general and holds
irrespective of whether the particles are fundamental or are low energy
excitations of an underlying condensed matter system (e.g. quasi-particles).

We begin with a Hilbert space which describes $n$ hard-core particles hopping 
on a $d$ dimensional lattice. The states can be labeled by listing the 
positions of the $n$ particles: $\ket{i_{1}, i_{2}, ... i_{n}}$.  The 
particles are identical so the states $\ket{i_{1}, i_{2}, ... i_{n}}$ do not 
depend on the order of $i_1,i_2,...,i_n$. For example,
\begin{displaymath}
 \ket{i_{1}, i_{2}, ... i_{n}}=\ket{i_{2}, i_{1}, ... i_{n}}
\end{displaymath}

A typical Hamiltonian for this system is of the form
\begin{displaymath}
H = \sum_{<ij>} (t_{ij} + t_{ji})
\end{displaymath}
where $t_{ij}$ are ``hopping operators'' with the property that
\begin{equation}
t_{ij} \ket{j, i_{1}, ..., i_{n-1}} \propto \ket{i, i_{1}, ..., i_{n-1}}
\label{hop}
\end{equation}
We assume that the hopping is local, \emph{i.e}
\begin{equation}
 [t_{ij}, t_{kl}]=0
\label{loc}
\end{equation}
if $i,j,k,l$ are all different.  

Our goal is to compute the statistics of the particles described by this
hopping Hamiltonian $H$. In the following, we show that the statistical angle
can be derived from the simple algebraic properties of the hopping operators.
More precisely, we show that the particles obey statistics
$e^{i\th}$ if
\begin{equation}
t_{il} t_{ki} t_{ij} = e^{i\th} t_{ij} t_{ki} t_{il}
\label{swap}
\end{equation}
for any three hopping operators $t_{ij}, t_{ki}, t_{il}$, where $j,k,l$ are
(distinct) neighbors of $i$ (ordered in the clockwise direction in the case of
$2$ dimensions).  The orientation convention in $2$ dimensions is necessary 
for the anyonic case.

The simplest examples where we can apply this formula are the cases of 
noninteracting hard-core bosons or fermions. In these cases the 
$n$ particle Hamiltonian can be written as
\begin{displaymath}
H = -t \sum_{<ij>} (c^{\dagger}_{i}c_{j} + c^{\dagger}_{j}c_{i})
\end{displaymath}
where the $c_{i}$'s are the boson or fermion annihilation operators. The
hopping operators are just $t_{ij} = -t c^{\dagger}_{i}c_{j}$. A little
algebra confirms that the boson and fermion hopping operators do indeed satisfy
\Eq{swap} with $e^{i\th} = +1$ and $-1$, respectively.    
 
We now give a general derivation of the formula. We begin with the state 
$\ket{i,j, ...}$ which contains particles at sites $i,j$ and other particles 
far away. 

\begin{figure}
\centerline{
\includegraphics[width=2.5in]{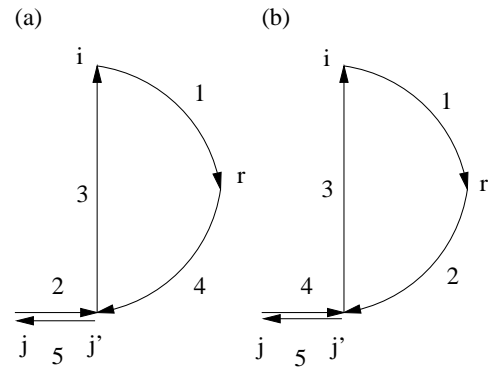}
}
\caption{ (a) The path of the particles in the construction of the swapped 
state $\ket{i,j,\text{swapped}}$. The numbers label the order in which the 
paths are traversed. (b) The path of the particles in the unswapped state 
$\ket{i,j,\text{unswapped}}$. Notice that the two states only differ in the 
ordering of the paths. However, in one case the particles switch places, while 
in the other, they do not.} 
\label{Stat}
\end{figure}

Imagine that we exchange the two particles at $i,j$ using an appropriate 
product of hopping operators. One expression for the final swapped state is
\begin{eqnarray*}
\ket{i,j, \text{swapped}} = (t_{jj'})(t_{j'p} ... t_{qr})(t_{il}
t_{lm} ... t_{nj'}) \\ (t_{j'j})(t_{rs} ... t_{ti}) \ket{i,j, ...} 
\end{eqnarray*}  
Here, we've grouped the hopping operators into 5 terms. Each of these terms 
moves a particle along some path on the lattice. The cumulative effect of all 5
terms is to exchange the particles (see Fig. \ref{Stat}a). 

Similarly, we can construct a path which doesn't swap the two particles. One
expression for the final unswapped state is
\begin{eqnarray*}
\ket{i,j, \text{unswapped}} = (t_{jj'})(t_{j'j})(t_{il}
t_{lm}  ... t_{nj'}) \\ (t_{j' p} ... t_{qr})(t_{rs} ... t_{ti}) \ket{i,j, ...}
\end{eqnarray*}
Once again, we've grouped the hopping operators into 5 terms. In this case, 
the cumulative effect of these terms is simply to move the two particles along 
independent loops without any swapping taking place (see Fig. \ref{Stat}b). 

Intuitively, we expect that the phase difference between the swapped and 
unswapped states is related to the statistical phase of the particles. In fact,
the special way we've constructed our states allows us to make a much stronger 
statement. Notice that the two states involve the same product of hopping 
operators. The only difference is the order of these operators. This means 
that, in the swapped and unswapped states, the two particles trace out the same
total path in configuration space. This is important because it implies that 
any phase which comes from gauge fields or other Berry phases contributes 
equally to the swapped and unswapped states. The phase difference between
the swapped and unswapped states is therefore exactly equal to the statistical 
phase of the particles.

This intuitive argument can be made rigorous using the discussion in the
previous section. Indeed, it's not hard to see that the two space-time paths 
traced out by the swapped and unswapped states are exactly the same locally - 
they only differ by a reordering. Furthermore, the phase difference between the
swapped and unswapped states can be written as a phase difference between two 
expressions of the form (\ref{prod}). Thus, taking the phase difference between
the swapped and unswapped states is completely equivalent to the procedure 
derived in the previous section.
 
We now compute this phase difference. We use the assumed algebraic relation
\begin{displaymath}
t_{j'p}t_{n j'} t_{j'j} = e^{i\th} t_{j'j} t_{n j'} t_{j'p}
\end{displaymath}
Applying this relation to the swapped state and reordering the hopping
operators using locality, we find
\begin{displaymath}
\ket{i,j,\text{swapped}} = e^{i\th} \ket{i,j,\text{unswapped}}
\end{displaymath}
The phase difference is $e^{i\th}$, so the particles obey statistics 
$e^{i\th}$, as claimed. A similar formula can be derived for the relative 
statistics of distinguishable particles. This formula, together with its 
derivation, is given in the appendix. (See \Eq{rel}).

\section{Fermions and Strings}
In this section we consider the properties of a lattice spin system with 
fermionic or anyonic excitations. We argue that the excitations are always 
created in pairs, and the creation operator for a pair of particles has a 
string-like structure, with the new particles located at the ends. One 
interpretation of this is that fermions never appear alone - they always come 
with some kind of gauge field.

We now make these statements more precise. Suppose we have a lattice spin 
system with fermionic excitations (the anyonic case is completely analogous). 
The total Hilbert space of the lattice spin model is a direct product of 
local Hilbert spaces $\mathcal{H}_{I}$, each associated with an 
individual lattice site $I$. We expect that the total Hilbert space contains a 
low energy subspace which is spanned by $n$ fermion states. The corresponding 
low energy effective Hamiltonian is given by
\begin{displaymath}
H_{\text{eff}} = PHP
\end{displaymath}
where $P$ is a projection operator onto the $n$ fermion subspace. Typically, 
$H_{\text{eff}}$ can be written as a sum
\begin{displaymath}
H_{\text{eff}} = \sum_{<ij>} (t_{ij} + t_{ji})
\end{displaymath}
where the $t_{ij}$ are hopping operators. We expect that the $t_{ij}$ are local
in the underlying spin degrees of freedom. That is, they act only within 
the local Hilbert spaces near sites $i$, $j$. \footnote{More 
precisely, the $t_{ij}$ are \emph{local at low energies}: the $t_{ij}$ can
be written as $t_{ij} = P\tilde{t}_{ij}P$ where $\tilde{t}_{ij}$ is local and 
commutes with the projection operator $P$.} We would like to point out that 
the lattice on which the fermion hops (labeled $i$) may not be the same as the
lattice formed by the local Hilbert spaces (labeled $I$).

Now, consider the string-like product of hopping operators
\begin{displaymath}
W_{ir} = t_{ij}t_{jk}t_{kl} ... t_{pq}t_{qr}
\end{displaymath}
This operator destroys a fermion at site $r$ and creates one at site $i$:
\begin{displaymath}
W_{ir} \ket{r, ...} \propto \ket{i, ...}
\end{displaymath}
where $\ket{r, ...}$, $\ket{i, ...}$ are low-energy states with fermions
at $r$, $i$, (and possibly other fermions far away). Furthermore, this operator
clearly has a string-like structure: it is made up of a string-like product of
operators, each of each is local in the underlying spin degrees of freedom.
The only issue is that this string operator might be trivial. That is,
$W_{ir}$ might act trivially on all the intermediate spins at $j, k, ..., q$, 
and only have a nontrivial effect near $i$ and $r$. In that case, $W_{ir}$ 
isn't really a string at all. 

This is a legitimate concern, since the string operator typically is trivial 
in the case of \emph{bosonic} quasi-particles. Consider, for example, a lattice
of noninteracting spins in a magnetic field: $H = -B \sum_{i} \si^{3}_{i}$. 
The ground state has $\si^{3}_{i} = 1$ for all $i$. The excitations are 
obtained by flipping one spin: $\si^{3}_{j} = -1$ for some $j$. If we perturb 
the Hamiltonian by a term $t \sum_{<ij>} \si^{1}_{i}\si^{1}_{j}$, these 
excitations acquire dynamics. The corresponding hopping operators are then 
$t_{ij} = \frac{t}{2} \si^{1}_{i}\si^{1}_{j}$ and the string operator is
\begin{eqnarray*}
W_{ir} &=& t_{ij}t_{jk} ... t_{qr} \\
       &\propto& (\si^{1}_{i}\si^{1}_{j})(\si^{1}_{j}\si^{1}_{k}) ... 
(\si^{1}_{q}\si^{1}_{r})\\
       &=& \si^{1}_{i}\si^{1}_{r}
\end{eqnarray*}
The operator $W_{ir}$ clearly creates a particle at position $i$ and
destroys one at position $r$. However, as we see above, it is trivial and has 
no string-like structure. The particle creation and annihilation operators
are completely local.

Our claim is that this can never happen in the case of emerging 
fermions. That is, the string operator $W_{ir}$ can never be written as a 
product
\begin{displaymath}
W_{ir} = A_{i}B_{r}
\end{displaymath}
where $A_{i}$, $B_{r}$ are operators which act only within the local Hilbert 
spaces near sites $i$, $r$. Physically, this means that the fermion creation 
and annihilation operators are never local in the underlying bosonic degrees 
of freedom. Instead, they naturally appear in pairs, with a string-like 
structure connecting them. \footnote{A more precise statement of our claim is 
that $W_{ir}$ cannot be written as a product $P_{s}A_{i}B_{r}P_{s}$ where 
$P_{s}$ is the projection operator onto the $n$ fermion subspace with no 
particles at any of the intermediate sites $j, k, ... , q$, and $A_{i}, B_{r}$ 
are local operators commuting with $P$. This statement is slightly stronger 
then the one made in the text, but the same proof applies.}

\emph{Proof:} We wish to use the algebraic formula (\ref{swap}) from the 
previous section. That formula relied on the locality of the hopping operators 
(\ref{loc}). Here we would like to use a weaker locality condition:
\begin{equation}
[t_{ij}, t_{kl}] = 0
\label{weakloc}
\end{equation}
if $i$, $j$, \emph{are far} from $k$, $l$. One can make a similar
argument using this weaker assumption. One arrives at a slightly weaker 
algebraic relation:
\begin{displaymath}
W_{il}W_{ki}W_{ij} = -W_{ij}W_{ki}W_{il}
\end{displaymath}
if $i$, $l$, $k$ are sufficiently far from each other. Since $W_{ki}W_{ij} =
W_{kj}$, we can rewrite this as
\begin{equation}
W_{il}W_{kj} = -W_{ij}W_{kl}
\label{fermi}
\end{equation}

Condition (\ref{fermi}) is a direct consequence of the fermionic nature 
of the quasi-particles. We will now show that it implies that the strings are 
nontrivial. Indeed, if the $W_{ir}$ could be written as $W_{ir} = A_{i}B_{r}$, 
then
\begin{eqnarray*}
W_{il}W_{kj} &=& A_{i}B_{l}A_{k}B_{j} \\
W_{ij}W_{kl} &=& A_{i}B_{j}A_{k}B_{l}
\end{eqnarray*}
But the $A$'s and $B$'s are local operators, so they commute with each other
when they are well separated. We can therefore rearrange the operators 
in these two equations to obtain
\begin{displaymath}
W_{il}W_{kj} = W_{ij}W_{kl}
\end{displaymath}
This directly contradicts the fermion condition (\ref{fermi}). We 
conclude that the string operator is always nontrivial. $\Box$

The presence of this nontrivial string indicates that fermions always appear 
together with some kind of gauge field. One way to see this is to consider
a closed loop of hopping operators $t_{ij}t_{jk} ... t_{pq}t_{qi}$. 
This string can be interpreted as a Wilson loop operator, since its phase
is precisely the accumulated phase of the particle when it traverses a loop.
The fact that it is nontrivial (that is, not equal to the identity operator)
means that the particle is coupled to a nontrivial gauge field.
                      
\section{A 2D example}

In this section, we present an exactly soluble lattice spin model with
fermionic excitations. The exactly soluble model provides a
concrete realization of the string picture discussed above.  The emerging
fermions turn out to be coupled to a $Z_{2}$ gauge field.

\begin{figure}
\centerline{
\includegraphics[width=2.1 in]{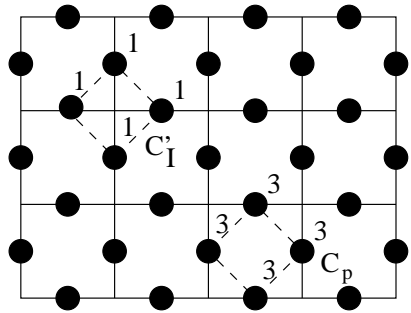}
}
\caption{
A schematic diagram depicting Kitaev's model. Each spin is drawn as a dot.  
The paths $C'_{\v I}$ and $C_{\v p}$ are drawn as dotted diamonds, and are 
labeled with $1$'s or $3$'s according to whether the corresponding term in
the Hamiltonian involves $\si^{1}$'s or $\si^{3}$'s.
}
\label{Slatt3}
\end{figure}

In this spin-1/2 system, proposed by Kitaev\cite{K9721}, the 
spins live on the \emph{links} of a square lattice (Fig. \ref{Slatt3}). The 
Hamiltonian is
\begin{equation}
H = -U \sum_{\v I} (\prod_{C'_{\v I}} \si^{1}_{\v j}) -
     g \sum_{\v p} (\prod_{C_{\v p}} \si^{3}_{\v j})
\label{Hkit}
\end{equation}
Here $\v i$ labels the links, $\v I$ labels the sites, and $\v p$ labels the 
plaquettes of the square lattice. Also, $C'_{\v I}$ denotes the loop 
connecting the four spins adjacent to site $\v I$, while $C_{\v p}$ denotes 
the loop connecting the four spins adjacent to plaquette $\v p$ 
(see Fig. \ref{Slatt3}).

This model is exactly soluble since all the terms in the Hamiltonian 
commute with each other. The ground state satisfies
\begin{displaymath}
\prod_{C'_{\v I}} \si^{1}_{\v j} = 1
\end{displaymath}
for all sites $I$, and
\begin{displaymath}
\prod_{C_{\v p}} \si^{3}_{\v j} = 1
\end{displaymath}
for all plaquettes $\v p$. There are two types of (localized) excited states.
We can have a site where
\begin{displaymath}
\prod_{C'_{\v I}} \si^{1}_{\v j} = -1
\end{displaymath}
or we can have a plaquette where
\begin{displaymath}
\prod_{C_{\v p}} \si^{3}_{\v j} = -1
\end{displaymath}
We call the first type of excitation a ``charge'' and the second
type of excitation a ``flux.'' Static charge and flux configurations are exact
eigenstates of the above Hamiltonian. Thus, the charge and flux quasi-particles
have no dynamics.

This lack of dynamics is a special feature of the above model. However, we are
interested in the properties of a \emph{generic} Hamiltonian in the same 
quantum phase. Thus, we need to perturb the system, and analyze the resulting 
dynamics. The simplest nontrivial perturbation is
\begin{equation}
H' = H + J_{1} \sum_{\v i} \si^{1}_{\v i} +J_{3} \sum_{\v i} \si^{3}_{\v i}
\label{h1}
\end{equation}
It's not hard to see that the first term allows the fluxes to hop from
plaquettes to adjacent plaquettes, while the second term allows the 
charges to hop from sites to neighboring sites.

We now calculate the statistics of the fluxes. To do this, we
restrict our Hamiltonian to the low energy subspace with $n$ fluxes and zero
charges. Within this subspace, our Hamiltonian reduces to
\begin{displaymath}
H'_{\text{eff}} = J_{1} \sum_{\v i} \si^{1}_{\v i}
\end{displaymath}

To make contact with our previous formalism, we write this as
\begin{displaymath}
H'_{\text{eff}} = \sum_{<\v p\v q>} (t_{\v p\v q} + t_{\v q\v p}) 
\end{displaymath}
where the sum is taken over adjacent plaquettes, $\v p,\v q$, $t_{\v p\v q}$ 
is defined by $t_{\v p\v q} = \frac{J_{1}}{2} \si^{1}_{\v i}$, and $\v i$ the
link joining $\v p$ and $\v q$.

To calculate the statistics, we need to compare $t_{\v p\v q}t_{\v r\v p}t_{\v
p\v s}$ with $t_{\v p\v s}t_{\v r\v p}t_{\v p\v q}$. Well, it's obvious from
the definition that all the hopping operators $t_{\v p\v q}$ commute with one
another. Therefore,
\begin{displaymath}
t_{\v p\v q}t_{\v r\v p}t_{\v p\v s} = t_{\v p\v s}t_{\v r\v p}t_{\v p\v q}
\end{displaymath}
We conclude that the fluxes are bosons. In the same way, one can show that the 
charges are also bosons.

Next, we consider the bound state of a flux and a charge. That is, we consider 
excitations with a flux through some plaquette $\v p$ and also a charge 
at one of the sites $\v I$ adjacent to $\v p$.

These bound states are not actually stable for the above Hamiltonian (\ref{h1})
- the charge and flux will separate from one another over time. However, one 
can imagine modifying the Hamiltonian so that charges and fluxes prefer to be
adjacent to each other. In this case, the bound state is a true quasi-particle.

\begin{figure}
\centerline{
\includegraphics[width=1.5 in]{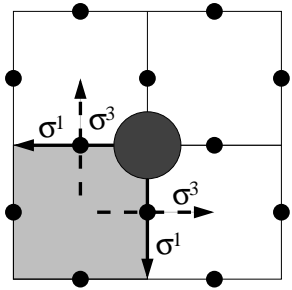}
}
\caption{
A bound state of a charge, denoted by a dark circle, and a flux, denoted by
a shaded square. The arrows illustrate the four ways the bound state can hop.
The two charge hops are denoted by solid arrows, and have hopping operator
$\si^{3}$. The flux hops are denoted by dotted arrows, and have hopping 
operator $\si^{1}$. Notice that each charge hopping operator anti-commutes
with a corresponding flux hopping operator, but everything else commutes. 
}
\label{Bstate}
\end{figure}

Suppose we've made such a modification. We can then consider the statistics of
the bound state. If we restrict our Hamiltonian (\ref{h1}) to the low energy
subspace with $n$ bound states, then our Hamiltonian reduces to
\begin{displaymath}
H'_{\text{eff}} = J_{1} \sum_{\v i} \si^{1}_{\v i} +
J_{3} \sum_{\v i} \si^{3}_{\v i}
\end{displaymath}
To understand the effect of these terms, imagine we have a bound state with a 
flux at $\v p$ and a charge at $\v I$. It's not hard to see that the first 
term allows the flux to hop to the two neighboring plaquettes which are also 
adjacent to $\v I$. Similarly, the second term allows the charge to hop to the 
two neighboring sites, adjacent to $\v p$. All other hopping destroys the 
bound state and is therefore forbidden for energetic reasons. Thus, the 
perturbation gives rise to $4$ types of hopping operators - $2$ corresponding 
to fluxes and $2$ corresponding to charges. (See Fig. \ref{Bstate}).

Formally, we can write our low-energy Hamiltonian as
\begin{displaymath}
H'_{\text{eff}} = \sum_{<(\v p,\v I)(\v q,\v J)>} (t_{(\v p,\v I)(\v q,\v J)} 
+ t_{(\v q,\v J)(\v p,\v I)})
\end{displaymath}
where the hopping operators are defined by
\begin{displaymath}
t_{(\v p,\v I)(\v q,\v J)} = \frac{J_{1}}{2} \si^{1}_{\v i} \text{ or } 
\frac{J_{3}}{2} \si^{3}_{\v i}
\end{displaymath}
depending on whether $(\v p,\v I), (\v q,\v J)$ differ by a flux hop or a 
charge hop. In the first case, $\v i$ is defined to be the link joining $\v p$ 
and $\v q$, while in the second case, $\v i$ is the link joining $\v I$ and 
$\v J$.

To calculate the statistics we need to compare a product of the form
$t_{1}t_{2}t_{3}$ with the product $t_{3}t_{2}t_{1}$, where the $t_{k}$ are
hopping operators involving a single bound state at position $(\v p,\v I)$.
(We could write out these expressions precisely, but it's messy and not very
enlightening). 

Now, as we discussed above, there are four different ways that a bound state 
at $(\v p,\v I)$ can hop - two charge hops, and two flux hops. Each of the flux
hopping operators can be paired with a corresponding charge operator which
involves the same link $\v i$. It's easy to see that each of the flux
operators anti-commutes with the corresponding charge operator (since one
involves a $\si^{1}$, while the other involves a $\si^{3}$). However, 
everything else commutes. (See Fig. \ref{Bstate}).

With these facts in mind, we can compare $t_{1}t_{2}t_{3}$ with 
$t_{3}t_{2}t_{1}$. By our discussion above, any set of $3$ hopping operators
involving $(\v p,\v I)$ must contain exactly $2$ which anti-commute. Thus,
exactly $2$ of $t_{1}, t_{2}, t_{3}$ anti-commute. This implies that

\begin{displaymath}
t_{1}t_{2}t_{3} = -t_{3}t_{2}t_{1}
\end{displaymath}
We conclude that the bound states are fermions. Of course, this result is not 
that surprising once we notice that the charges and fluxes have relative 
statistics $e^{i\phi_{\text{rel.}}} = -1$.

We can easily check that the charges and fluxes have relative statistics $-1$.
Normally, we would just use the formula (\ref{rel}) derived in the appendix. 
However, this relation assumes that the two types of particles both hop on the 
same lattice. In our case, the charges hop on the square lattice, while the 
fluxes hop on the dual lattice. 

It's not hard to see that in our case the formula (\ref{rel}) needs to be 
modified to
\begin{displaymath}
t_{\v I\v J}t_{\v p\v q} = e^{i\phi_{\text{rel.}}}t_{\v p\v q}t_{\v I\v J}
\end{displaymath}
where $\v I,\v J$ are neighboring sites, $\v p,\v q$ are neighboring 
plaquettes, and the links connecting $\v I,\v J$, and $\v p,\v q$ are the same.
From our previous calculations we know that 
$t_{\v p\v q} = \frac{J_{1}}{2}\si^{1}_{\v i}$, and 
$t_{\v I\v J} = \frac{J_{3}}{2}\si^{3}_{\v i}$. Therefore, $t_{\v p\v q}$, 
$t_{\v I\v J}$ anti-commute, and the relative statistics are 
$e^{i\phi_{\text{rel.}}} = -1$. 

We now see why the two particles were called ``charges'' and ``fluxes'' - 
they have the same statistics and relative statistics as $Z_{2}$ charges and 
fluxes. It turns out that this connection with $Z_{2}$ gauge theory extends 
beyond the low energy regime - in fact, all the way to the lattice scale. One 
can show that the Kitaev model is exactly equivalent to standard $Z_{2}$ gauge 
theory coupled to a $Z_{2}$ Higgs field.

We argued earlier that whenever fermions or anyons occur in a bosonic system,
they are always created in pairs, and the pair creation operator has a 
string-like structure. The above exactly soluble model (\ref{Hkit}) provides a 
good example of this phenomenon.

\begin{figure}
\centerline{
\includegraphics[width=2.0 in]{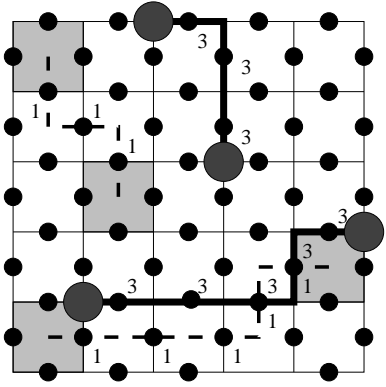}
}
\caption{
Three examples of string-like creation operators. The string
operators are drawn together with the particles they create 
at their endpoints. The thick solid line (drawn on the lattice) is an 
example of a charge string operator. The dotted line (drawn on the dual 
lattice) is an example of a flux string operator. At the bottom, we give an
example of a string operator for the bound state of a charge and flux. 
It is essentially a combination of a charge string and a flux string.
}
\label{String}
\end{figure}

We begin with the charges. We can construct the string operators 
associated with these particles by taking products of their hopping operators
along some path $P = \v{I_{1}}, ... \v{I_{n}}$ on the lattice. We find
\begin{eqnarray}
W(P) &=& t_{\v I_{1} \v I_{2}} t_{\v I_{2} \v I_{3}}  ... 
t_{\v I_{n-2} \v I_{n-1}} t_{\v I_{n-1} \v I_{n}}  \nonumber \\ 
&\propto& \si^{3}_{\v i_{1}} \si^{3}_{\v i_{2}}  ... \si^{3}_{\v i_{n-2}} 
\si^{3}_{\v i_{n-1}} 
\end{eqnarray}
where $\v i_{1}  ... \v i_{n-1} $ are the links along the path 
(see Fig. \ref{String}). Notice that $W(P)$ is nontrivial - it acts 
non-trivially on the spin degrees of freedom along the string $P$.

It is also a pair creation operator: If we apply $W(P)$ to the ground state,
the resulting state is an exact eigenstate with two charges - one located at 
each endpoint of $P$. One way to see this is to notice that $W(P)$ commutes 
with everything in the Hamiltonian except 
$\prod_{C'_{\v I_{1}}} \si^{1}_{\v j}$ and 
$\prod_{C'_{\v I_{n}}} \si^{1}_{\v j}$. The string anti-commutes with
these two operators, so when we apply it to the ground state, we
get a state with two charges located at $\v I_{1}, \v I_{n}$.

The case of the flux quasi-particles is very similar. We take the product of
the flux hopping operators along some path $P' = \v p_{1} , ... \v p_{n} $ on 
the dual lattice. We find
\begin{eqnarray}
W'(P') &=& t_{\v p_{1 }\v p_{2}} t_{\v p_{2} \v p_{3}}  ... 
t_{\v p_{n-2} \v p_{n-1}} t_{\v p_{n-1} \v p_{n}}  \nonumber \\ 
&\propto& \si^{1}_{\v i_{1}} \si^{1}_{\v i_{2}}  ... \si^{1}_{\v i_{n-2}} 
\si^{1}_{\v i_{n-1}} 
\end{eqnarray} 
where $\v i_{1}  ... \v i_{n-1} $ are the links along the path 
(see Fig. \ref{String}). Just as before, one can show that the string 
$W'(P')$ is a creation operator which creates two fluxes at the endpoints of 
$P'$.
 
Finally, consider the case of the bound state of the charge and the flux.
The hopping operator for the bound state is a combination of the charge and
flux hopping operators, so the string turns out to be a combination of the
charge and flux strings. Let ${P} = (\v p_{1}, \v I_{1}) ... (\v p_{n}, 
\v I_{n})$ be a path in bound state configuration space. Then the associated 
bound state string operator is 
\begin{equation}
{W}({P}) \propto \prod_{k} \si^{a_{k}}_{\v{i}_{k}} 
\end{equation}
where $\v i_{1}  ... \v i_{n-1 }$ are the links along the path and 
$a_{k} = 1$ or $3$ depending on whether $(\v p_{k}, \v I_{k})$ differ by
a flux hop or a charge hop, respectively (see Fig. \ref{String}). Once 
again, one can show that the string ${W}({P})$ is a creation operator for 
a pair of bound states located at the ends of ${P}$.

In each of these examples, the creation operator for fermions or anyons
has an extended string-like structure. It is important to note that
the position of this string is completely unobservable. That is, in each case,
the excited state $W(P)\ket{0}$ is independent of the position of $P$: it only
depends on the position of the \emph{endpoints} of $P$.

\section{Exactly soluble 3D model}

In two dimensional systems, one can create a fermion by binding a $Z_2$ vortex 
to a $Z_2$ charge.  This is how we obtained the fermion in the above spin-1/2 
model on the square lattice.  Since both the $Z_2$ vortex and $Z_2$ charge 
appear as the ends of open strings, the fermions also appear as the ends of 
strings. However, in three dimensions, we cannot change a boson into a fermion 
by attaching a $\pi$-flux. Thus one may wonder if fermions still appear as the 
ends of strings in (3+1)D. In this section, we study an exactly soluble 
spin-3/2 model on a cubic lattice.  We will show that the creation operator 
for fermions does indeed have a string-like structure. This example 
demonstrates that the string picture for fermions is more general then the 
flux-charge picture.

Our model has four states for each site of a cubic lattice. Thus we call it a 
spin-3/2 model. Let $\ga^{ab}$, $a,b \in \{x,\bar x, y,\bar y,z,\bar z \}$, 
$a \neq b$, be $4 \times 4$ hermitian matrices that satisfy
\begin{align}
\label{gabalg}
\ga^{ab}=&-\ga^{ba}=(\ga^{ab})^\dag  \nonumber\\
[\ga^{ab},\ga^{cd}]=& 0,\ \ \hbox{if $a,b,c,d$ are all different}
\nonumber\\
\ga^{ab}\ga^{bc} =& i \ga^{ac}, \ \ a\neq c
\nonumber\\
(\ga^{ab})^2 =& 1
\end{align}
A solution of the above algebra can be constructed by taking
pairwise products of Majorana fermion operators $\la^a$, 
$a \in \{x,\bar x, y,\bar y,z,\bar z \}$:
\begin{align}
 \ga^{ab}=& i\la^a\la^b  \nonumber\\
 \{ \la^a,\la^b\}= & 2\del_{ab}
\end{align}

The six Majorana fermion operators naturally require a space of dimension
$2^{6/2} = 8$, but if we restrict them to the space $\prod_{a} \la^{a} = 1$,
we obtain the desired $4 \times 4$ hermitian matrices. Alternatively, a more
concrete description of the $\ga^{ab}$ is given the appendix,
where we express the $\ga^{ab}$ in terms of Dirac matrices.  

In terms of $\ga^{ab}_{\v i}$, the exactly soluble spin-3/2 Hamiltonian can be 
written as
\begin{align}
\label{H3D}
H &=  -g \sum_{\v p} F_{\v p}
\end{align}
where $\v p$ labels all the square plaquettes in the cubic lattice, and
the $F_{\v p}$ are ``flux'' operators defined by
\begin{align}
 F_{\v p} =&
\ga^{yx}_{\v i} \ga^{\bar x y}_{\v i+\v{\hat{x}}} 
 \ga^{\bar y\bar x}_{\v{i} + \v{\hat{x}}+\v{\hat{y}}} 
\ga^{x\bar y}_{\v i+\v{\hat{y}}} ,
\nonumber\\
\text{or}\ &
\ga^{zy}_{\v i} \ga^{\bar y z}_{\v i+\v{\hat{y}}} 
 \ga^{\bar z\bar y}_{\v i+\v{\hat{y}}+\v{\hat{z}}} 
\ga^{y\bar z}_{\v i+\v{\hat{z}}} ,
\nonumber\\
\text{or}\  &
\ga^{xz}_{\v i} \ga^{\bar z x}_{\v i+\v{\hat{z}}} 
 \ga^{\bar x\bar z}_{\v i+\v{\hat{z}}+\v{\hat{x}}} 
\ga^{z\bar y}_{\v i+\v{\hat{x}}} .
\end{align}
depending on the orientation of the plaquette $\v{p}$.  Just as in the Kitaev
model,\cite{K9721} all the $F_{\v p}$ commute with each other and all the
$F_{\v p}$ have only two eigenvalues: $\pm 1$. Thus, we can solve the
Hamiltonian by simultaneously diagonalizing all the $F_{\v p}$.  If
$\ket{\{f_{\v p}\}}$ is a common eigenstate of $F_{\v p}$ with $F_{\v p}
\ket{\{f_{\v p}\}} = f_{\v p} \ket{\{f_{\v p}\}}$, $f_{\v p}=\pm 1$, then it
is also an energy eigenstate with energy
\begin{displaymath}
 E(\{f_{\v p}\}) = -g \sum_{\v p} f_{\v p}
\end{displaymath}
The one subtlety is that the $f_{\v p}$ are not all independent. If $C$ is the 
surface of a unit cube, then we have the operator identity
\begin{displaymath}
 \prod_{\v p\in C} F_{\v p}=1
\end{displaymath}
It follows that 
\begin{displaymath}
 \prod_{\v p\in C} f_{\v p}=1
\end{displaymath}
for all cubes $C$. This constraint means that the spectrum of our model is
identical to a $Z_{2}$ gauge theory on a cubic lattice. The ground state can
be thought of as a state with no flux: $f_{\v p}=1$ for all $\v p$. Similarly,
the elementary excitations are small flux loops where
\begin{displaymath}
f_{\v p} = -1
\end{displaymath}
for the four plaquettes $\v{p}$ adjacent to some link $\<\v{i}\v{j}\>$. We can 
think of these excitations as quasi-particles which live on the \emph{links} 
of the cubic lattice.

We would like to compute the statistics of these excitations. It is
tempting to assume that they are bosons since the model is almost the same as
a $Z_{2}$ gauge theory. However, as we will show, this superficial similarity 
is misleading: the flux loops are actually fermions.

\begin{figure}
\centerline{
\includegraphics[width=1.5in]{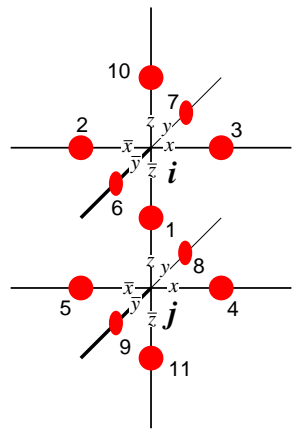}
}
\caption{
A fermion on link $1$ can hop onto 10 different links: $2$ - $11$.
The hopping $1\to 2$ is generated by $\ga^{\bar x\bar z}_{\v i}$, the
hopping $1\to 10$ by $\ga^{z \bar z}_{\v i}$, and the
hopping $1\to 4$ by $\ga^{xz}_{\v j}$.
}
\label{3Dhop}
\end{figure}

As in the Kitaev model, the quasi-particles are exact eigenstates of our
Hamiltonian. Thus, they have no dynamics and it is difficult to compute 
their statistics. However, this is a special feature of our model. We
need to perturb the theory to understand generic states in the same
quantum phase. The simplest perturbation is
\begin{displaymath}
H' = H + t \sum_{\v{i},a,b} \ga^{ab}_{\v{i}}
\end{displaymath}
To compute the statistics, we need to restrict ourselves to the low energy
subspace with $n$ quasi-particles. In this subspace, $H'$ reduces to
\begin{displaymath}
H'_{\text{eff}} =  t \sum_{\v{i},a,b} \ga^{ab}_{\v{i}}
\end{displaymath}
The effect of each term $\ga^{ab}_{\v{i}}$ is to allow the quasi-particles
to hop between the two links $\<\v{i}(\v i+\v{\hat{a}})\>$ and 
$\<\v{i}(\v i+ \v{\hat{b}})\>$ adjacent to site $\v{i}$. 
Here $\v{\hat a}=\v{\hat{x}}$ if $a=x$, $\v{\hat a}=-\v{\hat{x}}$ if 
$a=\bar x$, \etc. (See Fig. \ref{3Dhop}.) Thus, our Hamiltonian 
can be written in the standard hopping form:
\begin{displaymath}
H'_{\text{eff}} = 
\sum_{\v{i},a,b} (t_{\<\v{i}(\v i+ \v{\hat{a}})\>\<\v{i}(\v i+ \v{\hat{b}})\>}
+ t_{\<\v{i}(\v i+ \v{\hat{b}})\>\<\v{i}(\v i+ \v{\hat{a}})\>})
\end{displaymath}
where $t_{\<\v{i}(\v i+ \v{\hat{a}})\>\<\v{i}(\v i+ \v{\hat{b}})\>}= 
\frac{t}{2} \ga^{ab}_{\v{i}}$.

To calculate the statistics we need to compare a product of the form
$t_{1}t_{2}t_{3}$ with the product $t_{3}t_{2}t_{1}$, where the $t_{k}$ are
hopping operators involving a single link $\<\v{i}\v{j}\>$. (As in the Kitaev
model, we could write out these expressions explicitly, but it's not very
enlightening).

Note that a quasi-particle on the link $\<\v{i}\v{j}\>$ can hop to any of the 
$10$ neighboring links, $5$ of which are adjacent to site $\v{i}$, and $5$ 
of which are adjacent to site $\v{j}$ (see Fig. \ref{3Dhop}). Using the
algebra (\ref{gabalg}) or the Majorana fermion representation, we find that
the $5$ hopping operators associated with site $\v{i}$ all anti-commute with
each other and similarly for site $\v{j}$. On the other hand, each of 
the operators associated with $\v{i}$ commute with each of the operators
associated with $\v{j}$.

With these facts in mind, we can now compare $t_{1}t_{2}t_{3}$ with
$t_{3}t_{2}t_{1}$. There are essentially two cases: either all three of the
$t_{k}$'s are associated with a single site $\v{i}$ or $\v{j}$, or two involve
one site and one involves the other. In the first case, all the $t_{k}$'s
anti-commute; in the second case, a pair of the $t_{k}$'s anti-commute and
everything else commutes. In either case, we have
\begin{displaymath}
t_{1}t_{2}t_{3} = -t_{3}t_{2}t_{1}
\end{displaymath} 
We conclude that the quasi-particles are indeed fermions.

\begin{figure}
\centerline{
\includegraphics[width=2.6in]{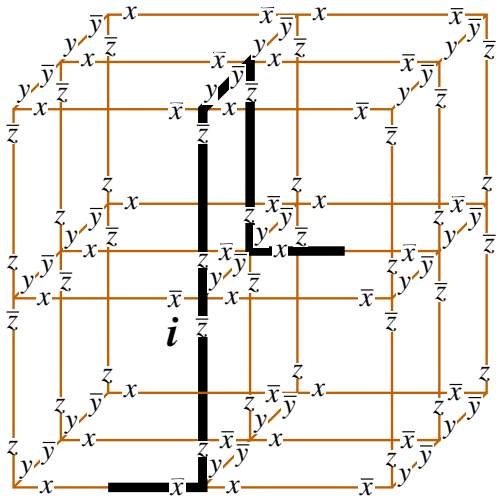}
}
\caption{
An example of a string in the 3D model. This string corresponds to the 
operator $W(P) \propto \ga_{\v{i}-\v{\hat{z}}}^{\bar x z} 
\ga_{\v{i}}^{\bar z z} \ga_{\v{i}+\v{\hat{z}}}^{\bar z y} 
\ga_{\v i+\v{\hat{z}}+\v{\hat{y}}}^{\bar y \bar z} 
\ga_{\v i+\v{\hat{y}}}^{zx}$.
}
\label{3Dlatt}
\end{figure}

Next, we construct the associated string operator. Taking a product of 
hopping operators along a path (of links),
$P = \<\v i_1 \v i_2 \>,\<\v i_2 \v i_3 \>, ... \<\v i_{n-1} \v i_{n} \>$,  
gives
\begin{equation}
\label{3Dstring}
W(P) \propto \ga_{\v i_1 }^{a_1b_1}\ga_{\v i_2 }^{a_2b_2}...
\ga_{\v i_n }^{a_nb_n}
\end{equation}
where $b_m,a_{m+1}$ is the pair of indices associated with the link 
$\<\v i_m \v i_{m+1} \>$. (See Fig.\ref{3Dlatt}.) Notice that the string 
operator commutes with the Hamiltonian, except at the ends. Thus, when we 
apply $W(P)$ to the ground state, the only effect is to flip the signs of 
$f_{\v p}$ for the plaquettes $\v p$ adjacent to the links 
$\<\v i_1 \v i_2\>$, $\<\v i_{n-1} \v i_{n}\>$. This means that the open 
string operator creates a pair of fermions at its two ends. 

\section{Conclusion}

In this paper we have derived a general relation (\ref{swap}) between the 
statistics of a lattice particle and the algebra of its hopping operators. This
relation allows us to analyze emerging fermions in three dimensions. We have 
also shown that there is a close connection between strings and fermions. The 
statistical algebra (\ref{swap}) is fundamental to this connection since it 
allows us to determine the statistics of the ends of strings from the 
structure of the string operator. 

It is interesting to put this string picture of emerging fermions in context. 
Indeed, it is well-known that gauge theories and strings are closely 
related - the strings correspond to electric flux lines in the associated gauge
theory. \cite{W7159,W7445,KS7595,BMK7793,P7947,P981,Walight} Thus, it appears 
that the fundamental concepts of Fermi statistics, gauge theory, and 
strings are all connected. They are all just different aspects of a 
kind of order - topological order - in bosonic lattice systems.

This research is supported by NSF Grant No. DMR--01--23156
and by NSF-MRSEC Grant No. DMR--02--13282.

\appendix               
\section{Relative statistics}
In this section, we derive an algebraic formula, analogous to (\ref{swap}), 
for the relative statistics of distinguishable particles. We begin with a 
Hilbert space which describes two types of hard-core particles hopping on a 2D 
lattice. For concreteness, say that there are $m$ particles of type $1$ and 
$n$ particles of type $2$. The states can be labeled by listing the positions 
of the two types of particles: $\ket{i_{1}, ... , i_{m} ; j_{1}, ... , j_{n}}$.

A typical Hamiltonian for this system is of the form
\begin{displaymath}
H = \sum_{<ij>} (t^{1}_{ij} + t^{1}_{ji} + t^{2}_{ij} + t^{2}_{ji})
\end{displaymath}
where $t^{1}_{ij}, t^{2}_{ij}$ are hopping operators for the two types of
particles.

We wish to calculate the relative statistics of the two types of particles.
As before, the statistics are related to the simple algebraic properties of
the hopping operators. Specifically, we will show that the particles have 
relative statistics $e^{i\phi}$ if
\begin{equation}
(t^{2}_{ip}t^{2}_{pj})(t^{1}_{kp}t^{1}_{pl}) = 
e^{i\phi}(t^{1}_{kp}t^{1}_{pl})(t^{2}_{ip}t^{2}_{pj})
\label{rel}
\end{equation}
Here, $i,k,j,l$ are (distinct) neighbors of $p$ oriented in the clockwise 
direction.

\begin{figure}
\centerline{
\includegraphics[width=2.5in]{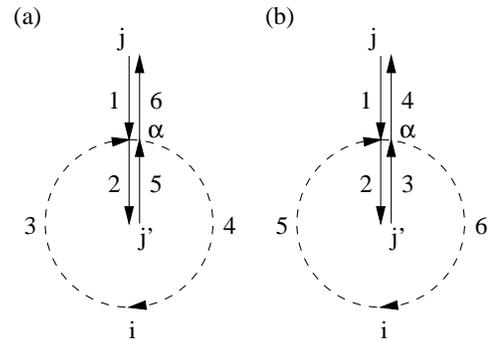}
}
\caption{ (a) The path of the particles in the construction of the wound state 
$\ket{i;j,\text{wound}}$. The dotted line is the path of particle $1$, and
the solid line is the path of particle $2$. The numbers label the order in 
which the paths are traversed. (b) The path of the particles in the unwound 
state $\ket{i;j,\text{unwound}}$. Notice that the two states only differ in 
the ordering of the paths. However, in one case the particles wind around each 
other, while in the other, they do not.} 
\label{Relstat}
\end{figure}

We begin with the state $\ket{i;j ...}$ which contains a type $1$ particle at 
$i$, a type $2$ particle at $j$, and other particles which are far away.

Imagine that we wind particle $1$ around particle $2$ using an 
appropriate product of hopping operators. One expression for the final 
wound state is:
\begin{eqnarray*}
\ket{i;j, \text{wound}} = (t^{2}_{j \alpha})(t^{2}_{\alpha j'}) 
(t^{1}_{il}...t^{1}_{m\alpha}) \\ (t^{1}_{\alpha n}...
t^{1}_{pq}t^{1}_{qi})(t^{2}_{j'\alpha})(t^{2}_{\alpha j})  
\ket{i;j ...}
\end{eqnarray*}
Here, we've grouped the hopping operators into 6 terms. The cumulative effect 
of these terms is to wind one particle around the other (see 
Fig. \ref{Relstat}a).
 
Similarly, we can construct a state where the two particles don't wind
around each other (see Fig. \ref{Relstat}b):
\begin{eqnarray*}
\ket{i;j, \text{unwound}} = (t^{1}_{il}...t^{1}_{m\alpha})
(t^{1}_{\alpha n}...t^{1}_{pq}t^{1}_{qi})(t^{2}_{j \alpha}) \\ 
(t^{2}_{\alpha j'})(t^{2}_{j'\alpha})(t^{2}_{\alpha j})\ket{i;j ...}
\end{eqnarray*}

As before, the two states involve the same total path in configuration space. 
Thus, the phase difference between the two states is precisely the relative 
statistics of the particles. This can be made rigorous using an argument 
similar to the exchange statistics case.

We now calculate this phase difference. We use the assumed algebraic relation
\begin{displaymath}
(t^{2}_{j\alpha}t^{2}_{\alpha j'})(t^{1}_{m\alpha}t^{1}_{\alpha n}) =
e^{i\phi}(t^{1}_{m\alpha}t^{1}_{\alpha n})(t^{2}_{j\alpha}t^{2}_{\alpha j'})
\end{displaymath}
Applying this relation to the wound state, and reordering the hopping operators
using locality, we find
\begin{displaymath}
\ket{i;j, \text{wound}} = e^{i\phi} \ket{i;j, \text{unwound}}
\end{displaymath}
This establishes the desired result: the particles have relative statistics 
$e^{i\phi}$. 

A good consistency check for this formula can be obtained by considering the 
relative statistics of \emph{two particles of the same type}. In this case,
we expect the angle for the relative statistics to be exactly twice the 
angle for the exchange statistics: 
\begin{displaymath}
e^{i\phi_{\text{rel.}}} = e^{2i\th_{\text{stat.}}}
\end{displaymath}
One way to see this is that exchanging the particles twice in the same
direction is topologically equivalent to winding one particle around the
other.

This result, which has a topological character to it, can be derived 
algebraically from our two formulas. We start with the expression
$(t_{ip}t_{pj})(t_{kp}t_{pl})$. Applying the exchange
statistics formula \Eq{swap}, we find:
\begin{eqnarray*}
(t_{ip}t_{pj})(t_{kp}t_{pl}) &=& t_{ip}(t_{pj}t_{kp}t_{pl}) \\
                          &=& t_{ip}(e^{i\th_{\text{stat.}}}t_{pl}t_{kp}t_{pj})
\end{eqnarray*}
Applying the formula again gives
\begin{eqnarray*}
t_{ip}(t_{pl}t_{kp}t_{pj}) &=& (t_{ip}t_{pl}t_{kp})t_{pj} \\
                         &=& (e^{i\th_{\text{stat.}}}t_{kp}t_{pl}t_{ip})t_{pj}
\end{eqnarray*}
Combining these two equations gives
\begin{displaymath}
(t_{ip}t_{pj})(t_{kp}t_{pl}) = 
e^{2i\th_{\text{stat.}}}(t_{kp}t_{pl})(t_{ip}t_{pj})
\end{displaymath}
Comparing this with the relative statistics formula \Eq{rel}, we see that
$e^{i\phi_{\text{rel.}}} = e^{2i\th_{\text{stat.}}}$, as claimed.
 
\section{Dirac matrix representation of $\ga^{ab}$}
In the following, we represent $\ga^{ab}$'s in terms of the Dirac
matrices.  Note that \Eq{gabalg} implies that
\begin{displaymath}
 \{\ga^{ac},\ga^{bc}\}=2\del_{ab}
\end{displaymath}
Thus 
$\ga^{xz}$, $\ga^{\bar xz}$, $\ga^{yz}$, and $\ga^{\bar yz}$ satisfy the 
algebra of Dirac matrices. Introducing four Dirac matrices
\begin{align*}
\ga^{x} = & \si^1\otimes \si^1 , &
\ga^{\bar x} = & \si^2\otimes \si^1   \nonumber\\
\ga^{y} = & \si^3\otimes \si^1 , & 
\ga^{\bar y} = & \si^0\otimes \si^2  ,
\end{align*}
we can write 
\begin{equation}
 \ga^{az}=\ga^a,\ \ a=x,\bar x,y,\bar y.
\end{equation}
Similarly, $\ga^{x\bar z}$, 
$\ga^{\bar x\bar z}$, 
$\ga^{y\bar z}$, and
$\ga^{\bar y\bar z}$ satisfy the algebra of Dirac matrices.
We can express those operators as
\begin{equation}
 \ga^{a\bar z}=i\ga^a\ga^5,\ \ a=x,\bar x,y,\bar y.
\end{equation}
where $\ga^5=\ga^{x}\ga^{\bar x}\ga^{y}\ga^{\bar y}$.
Finally, for $a,b=x,\bar x,y,\bar y$, we have
\begin{equation}
 \ga^{ab}=i\ga^a\ga^b
\end{equation}
In this way, we can express all the $\ga^{ab}$, 
$a,b \in \{x,\bar x, y,\bar y,z,\bar z \}$, $a \neq b$ in terms of Dirac 
matrices.


\end{document}